\documentclass[a4paper, reprint, superscriptaddress,floatfix]{revtex4-1}

\newcommand\temptitle{Same and interconvertible high-pressure ice phases}

\usepackage{float}
\usepackage{graphicx}
\usepackage{xcolor}
\usepackage[version=4]{mhchem}

\usepackage[T1]{fontenc}  %
\usepackage{newtxtext}
\usepackage[slantedGreek]{newtxmath}

\usepackage[pdftex, bookmarks, pdffitwindow=false, pdfstartview=FitH, pdfdisplaydoctitle, colorlinks, plainpages=false, pdftitle={\temptitle},pdfauthor={Aleks Reinhardt, Mandy Bethkenhagen, Federica Coppari, Marius Millot, Sebastien Hamel, Bingqing Cheng}, pdfpagelabels, hypertexnames, citecolor={blue!50!black},linkcolor={blue!50!black}, urlcolor={blue!50!black}, pdflang={en}, hyperfootnotes=false, breaklinks]{hyperref}

\usepackage{siunitx}
\DeclareSIUnit\angstrom{\text{Å}}
\DeclareSIUnit\bar{bar}
\usepackage[style=iso]{datetime2}

\makeatletter
\renewcommand{\@seccntformat}[1]{}
\makeatother

\newcommand{\avg}[1]{\ensuremath{\left\langle#1\right\rangle}}
\newcommand{\tm}{\ensuremath{T_{\text{m}}}} 
\newcommand{\ficm}{\ensuremath{\Tilde{m}}} 

\newcommand{\Pbcm}{P\textit{bcm}}
\newcommand{\Pbca}{P\textit{bca}}
\newcommand{\PtwoOne}{P2$_1$2$_1$2}
\newcommand{\PfourTwo}{P4$_2$2$_1$2}
\newcommand{\sevenP}{VII$^\prime${}}
\newcommand{\sevenPp}{VII$^{\prime\prime}${}}

\newcommand{\figLabel}[1]{\textbf{\textsf{\MakeLowercase{#1}}}}  
\newcommand{\figLabelCapt}[1]{\textbf{\textsf{\MakeLowercase{#1}}}}  

\newcommand{\refSub}[2]{\hyperref[#2]{\ref{#2}\figLabelCapt{#1}}}
\newcommand{\figref}[1]{Fig.~\ref{#1}}
\newcommand{\figrefsub}[2]{Fig.~\refSub{#2}{#1}}
\newcommand{\secref}[2]{\hyperref[#2]{#1}}
\newcommand{\figrefSI}[1]{Supplementary Fig.~\ref*{#1}}

 \newcommand{\subfigimg}[3][,]{%
   \setbox1=\hbox{\includegraphics[#1]{#3}}
   \leavevmode\rlap{\usebox1}
   \rlap{\hspace*{0pt}\raisebox{\dimexpr\ht1-2\baselineskip}{#2}}
   \phantom{\usebox1}
 }

\begin{document}

\title{\temptitle}

\author{Aleks Reinhardt}
\affiliation{Yusuf Hamied Department of Chemistry, University of Cambridge, Lensfield Road, Cambridge, CB2 1EW, United Kingdom}

\author{Mandy Bethkenhagen}
\affiliation{\'Ecole Normale Sup\'erieure de Lyon,  Universit\'e Lyon 1, Laboratoire de G\'eologie de Lyon, CNRS UMR 5276, 69364 Lyon Cedex 07, France}

\author{Federica Coppari}
\affiliation{Lawrence Livermore National Laboratory, Livermore, California 94550, USA}

\author{Marius Millot}
\affiliation{Lawrence Livermore National Laboratory, Livermore, California 94550, USA}

\author{Sebastien Hamel}
\affiliation{Lawrence Livermore National Laboratory, Livermore, California 94550, USA}

\author{Bingqing Cheng}
\email{bingqing.cheng@ist.ac.at}
\affiliation{Institute of Science and Technology Austria, Am Campus 1, 3400 Klosterneuburg, Austria}

\date{\today}

\begin{abstract}
Most experimentally known high-pressure ice phases have a body-centred cubic (bcc) oxygen lattice.
Our atomistic simulations show that, amongst these bcc ice phases, ices VII, \sevenP{} and X are the same thermodynamic phase under different conditions, whereas superionic ice \sevenPp{} has a first-order phase boundary with ice \sevenP{}.
Moreover, at about \SI{300}{\giga\pascal}, the transformation between ice X and the \Pbcm{} phase entails a structural change but no apparent activation barrier,
whilst at higher pressures the barrier gradually increases.
Our study thus clarifies the phase behaviour of the high-pressure insulating ices and reveals peculiar solid--solid transition mechanisms not known in other systems.
\end{abstract}

\maketitle

\section{Introduction}
Water ice exhibits remarkable structural diversity: there are currently 20 experimentally confirmed polymorphs with distinct atomic arrangements~\cite{Salzmann2019, Hansen2021}.
Intriguingly, at high pressures between \SIrange{2}{200}{\giga\pascal},
with the exception of one face-centred cubic (fcc) superionic phase (ice~XVIII)~\cite{Millot2019},
all known phases of ices have a body-centred cubic (bcc) lattice of oxygen atoms~\cite{Loubeyre1999},
with subtle differences in the positions and the dynamics of the hydrogen atoms.

Ice~VII is the proton-disordered counterpart of the antiferroelectric phase ice~VIII, and hydrogen atoms can occupy any of the half-diagonals of the bcc unit cell.
Ice~\sevenP{} forms upon compression (above $\sim$\SI{40}{\giga\pascal} at \SI{300}{\kelvin}) during which hydrogen atoms exhibit substantial translational movements along the bcc half-diagonal, leading to a bimodal distribution in the hydrogen positions between two oxygen atoms.
Ice~\sevenPp{} has a stronger delocalization of protons compared to \sevenP{}, and is the solid phase that coexists with the liquid above $\sim$\SI{1000}{\kelvin}.
Above $\sim$\SI{80}{\giga\pascal}, the bimodal distribution of hydrogen positions becomes unimodal and peaks at the centre: this is ice~X.
Between $\sim$\SI{300}{\giga\pascal} and \SI{700}{\giga\pascal}, density functional theory (DFT) calculations based on the PBE functional predict the \Pbcm{} phase, with a distorted hexagonal-close-packed oxygen lattice, to be stable~\cite{Benoit1996}.
It has been proposed that \Pbcm{} forms due to a dynamic instability in ice X above $\sim$\SI{400}{\giga\pascal}~\cite{Caracas2008}.

We have only rather limited understanding of the thermodynamic distinctions and the transitions between these high-pressure ice phases.
For example, it is not fully resolved whether some of these phase transitions are of first order.
The problem is partly due to the limited resolution of the experimental measurements at high pressures,
and partly because that system size and time scale accessible to DFT molecular-dynamics (MD) simulations are far from what is required to approach the thermodynamic limit of ice systems.
Hernandez and Caracas~\cite{Hernandez2016} found a first-order transition between \sevenP{} and \sevenPp{} using DFT MD,
based on the sudden changes in the diffusivity, the delocalization of the H atoms, internal energy and elastic constants.
The prediction of this isostructual first-order transition was later corroborated by the experimental evidence of a kink on the melting line $\tm$ at \SI{14.6}{\giga\pascal}, which has been interpreted to correspond to the \sevenP{}--\sevenPp{} transition line meeting $\tm$~\cite{Queyroux2020}.
By contrast, for the VII--\sevenP{}--X transition sequence, experiments did not find any first-order features in the equations of state~\cite{Hemley1987,Wolanin1997,Loubeyre1999, Mendez2021}.
Similarly, a VII/X equation of state parameterized from DFT-MD simulations suggests no first-order transition occurs, but does not exclude a possible second-order transition~\cite{French2015}, and effective potentials for the energy as a function of the proton position between oxygen atoms, constructed from DFT simulations, have been shown to convert continuously from a double well (ice VII) to a single well (ice X)~\cite{Trybel2020}.
Some recent work proposed a theoretical framework based on dynamic partition functions to gain insight into the transition between ice VII and its analogues with faster proton dynamics~\cite{Ye2021}.
In DFT simulations, X was observed to transform continuously into \Pbcm{} upon increased pressure at \SI{0}{\kelvin} without a noticeable barrier starting at about \SI{350}{\giga\pascal} via a transformation path analogous to that proposed for the bcc--hcp transitions in many metals~\cite{Benoit1996}.

From a thermodynamic point of view, it is thus not clear which of these ice phases are actually different phases of water 
(i.e.~separated by first-order phase boundaries, with chemical potentials crossing at coexistence points),
rather than the same thermodynamic phase exhibiting different behaviours at different conditions.
Determining the nature of these phase boundaries in theoretical calculations entails looking for signs of hysteresis or discontinuities in density or energy during phase transitions, or computing the form of the chemical potentials of the phases.
Both approaches require long simulations of large system sizes in order to approach the thermodynamic limit, and consequently standard DFT simulations are generally not feasible because of their severe system size and timescale limitations.
Furthermore, the transition pathways amongst the high-pressure ice phases remain largely elusive.
Understanding how such transitions occur and in particular the magnitude of the kinetic barrier provides additional insights on the structural relationships between the ice phases,
and is also crucial for designing and interpreting high-pressure experiments~\cite{Millot2018,Millot2019,Prakapenka2021,Loubeyre1999,Queyroux2020}.

In this study, we use a first-principles description of high-pressure water based on the Perdew--Burke--Ernzerhof (PBE)~\cite{Perdew1996} approximation to the exchange-correlation functional,
as well as a recently constructed machine-learning potential (MLP)~\cite{Cheng2021} fitted to the PBE reference that has been employed to predict the phase behaviour of superionic water.
We first compute the phase diagram of ice at $P<\SI{100}{\giga\pascal}$ for the MLP
and probe the nature of the phase boundaries between all the ice phases that share the bcc oxygen lattice,
and then investigate the X--\Pbcm{} solid--solid transition mechanism at higher pressures.
Crucially, the combination of first-principles methods, MLP, free-energy estimations and enhanced sampling methods enables us to provide both a thermodynamic picture of the high-pressure ice phases
and a detailed mechanistic view for the transitions between them.
We not only account for the influence of thermal and quantum fluctuations on the free energies, but also model the dynamic transitions between the ice phases.

\section{bcc ices}

Between \SI{2}{\giga\pascal} and \SI{200}{\giga\pascal} and below the melting line, 
except for ice~XVIII~\cite{Millot2019},
all the relevant ice phases (VII, \sevenP{}, \sevenPp{}, VIII, X) have a bcc lattice of oxygen atoms and differ by the position and the dynamics of the hydrogen atoms.
In this section, we investigate whether there are structural differences between these bcc-based ices or if any differences arise entirely from the dynamics of the H atoms, and whether phase transitions between them are of first order.
Unlike previous studies, here we use a thermodynamic approach by focusing on chemical potentials of sufficiently large systems rather than solely relying on the equations of states computed from DFT.

\begin{figure*}
   \centering
  \includegraphics[]{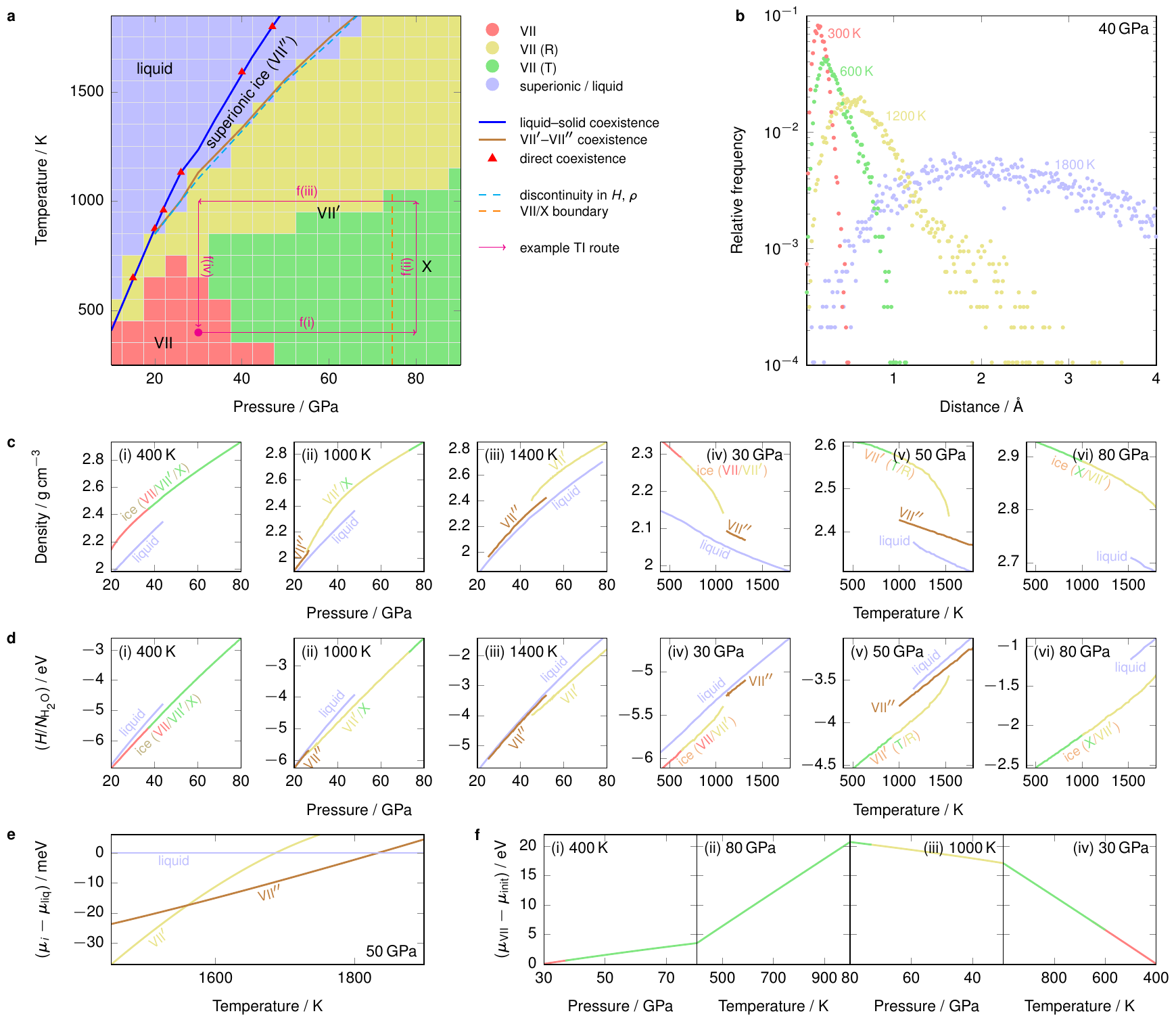}
   \caption{\textbf{Low-pressure bcc phase behaviour.}
   \figLabelCapt{a} Low-pressure phase diagram. 
   The background colour corresponds to the classification shown in panel~\figLabelCapt{b}. Blue and brown solid coexistence lines were computed from chemical-potential differences using thermodynamic integration (TI). 
   Several points were benchmarked by direct-coexistence simulations at the points indicated. 
   The dashed cyan line corresponds to points where the enthalpy and density are discontinuous along isobars (see panels~\figLabelCapt{c}--\figLabelCapt{d}).
   The dashed orange line corresponds to the approximate point where the first secondary maximum in the O--H pair correlation function disappears.
   The thin magenta line gives an example TI route discussed in the main text (see panel~\figLabelCapt{f}).
  \figLabelCapt{b} Mean frequency of proton displacement after \SI{0.4}{\pico\second} for MD simulations started from bcc ice at \SI{40}{\giga\pascal}. 
  Red corresponds to a symmetric unimodal distribution at small displacement (`static ice'). 
  Green corresponds to distributions with a mode at small displacement and a significant skewness to the right (`ice VII (T)').
  Yellow corresponds to a distribution with a mode at medium displacement and a skewness to the right (`ice VII (R)').
  Blue corresponds to a unimodal distribution at a large displacement.
  This classification of phases follows Ref.~\citenum{Zhuang2020}.     
   \figLabelCapt{c} Density and \figLabelCapt{d} enthalpy of the phases indicated along selected isotherms [(i)--(iii)] and isobars [(iv)--(vi)].
   The changeover in colour for ices VII/\sevenP{}/X is approximate and based on the classification of panel~\figLabelCapt{b}.
   \figLabelCapt{e} Example discontinuity in the gradients of the chemical potentials for the phase transitions involving the liquid, \sevenP{} and \sevenPp{} at \SI{50}{\giga\pascal}.
   \figLabelCapt{f} Change in chemical potential along the example TI route illustrated in panel~\figLabelCapt{a}.
   }
   \label{fig:lowT-classification}
\end{figure*}

The thermodynamic approach is made feasible by employing the MLP, which has first-principles accuracy but with orders-of-magnitude lower cost.
To validate the MLP, we first compare the lattice constants, potential energy and diffusivities computed from the MLP MD simulations with previous first-principles molecular-dynamics (FPMD) simulations~\cite{Hernandez2018,Queyroux2020} [\figrefSI{fig:X-EOS}], which suggests that the MLP is able to capture the dynamic and EOS differences between the different bcc ice phases.
Based on the data obtained from brute-force simulation using the MLP, we first make an initial classification of the ice phases, in order to compare with the current consensus about the ice phase diagram.
In \figrefsub{fig:lowT-classification}{a}, we show the low-pressure phase behaviour of the system alongside a line indicating where a discontinuity in density and enthalpy occurs in brute-force simulations along isobars.
Beyond $\sim$\SI{40}{\giga\pascal}, this line is consistent with the threshold of proton diffusivity ($D_\text{H}>\SI{e-8}{\metre\squared\per\second}$) used as a proxy for the superionic transition by Cheng~\textit{et al.}~\cite{Cheng2021}.
The background colour in \figrefsub{fig:lowT-classification}{a} corresponds to the dominant dynamic phase in brute-force MD simulations started from ice X and equilibrated over long times, using a classification introduced by Zhuang~\textit{et al.}~\cite{Zhuang2020, Ye2021}: we classify phases according to the frequency of displacements of protons after \SI{0.4}{\pico\second} by considering where the mode of the displacement distribution is and the skewness of the distribution [\figrefsub{fig:lowT-classification}{b}].
This classification enables a qualitative distinction to be made in the behaviour of ice VII and the dynamic nature of the proton diffusion in ice~\sevenP, which we have classed as rotational (`R') and translational (`T') for consistency with Ref.~\citenum{Zhuang2020}.
We also provide in \figrefsub{fig:lowT-classification}{a} an approximate~\cite{Hernandez2018} line beyond which the protons exhibit only one peak in the O--H pair correlation function at distances smaller than the nearest O--O distance, which serves to delineate ice X from ice VII.

The qualitative phase behaviour we observe is rather similar to that studied by Zhuang~\textit{et al.}~\cite{Zhuang2020,Ye2021} and Queyroux~\textit{et al.}~\cite{Queyroux2020, Hernandez2018}.
The presence of an apparent discontinuity in density and enthalpy in the superionic transition may suggest that this is a first-order transition, although it is often in practice difficult to distinguish between a discontinuous and merely a rapidly changing function.
Moreover, for the remaining phases, based on dynamic criteria alone it is not immediately obvious whether they are thermodynamically distinct.
To ascertain the nature of the phase transitions in this relatively low-pressure region of the phase diagram, we computed the liquid--solid coexistence curve using thermodynamic integration along isotherms and isobars~\cite{Reinhardt2021, Reinhardt2020, Vega2008} of the appropriate solid and liquid phases starting from an initial coexistence point determined by direct coexistence, and verified by direct-coexistence simulations~\cite{Opitz1974} at several other points.
We then determined the \sevenP{}--\sevenPp{} coexistence line by thermodynamic integration along isobars starting from the liquid--solid coexistence curve.
The resulting thermodynamic coexistence curve [\figrefsub{fig:lowT-classification}{a}, brown line] largely follows the locus of apparent discontinuities of the enthalpy and density in brute-force simulations; however, despite a relatively low degree of hysteresis, the gradients of the chemical potentials of the two phases of ice are different at either side of the coexistence point [\figrefsub{fig:lowT-classification}{e}], confirming that this is indeed a first-order phase transition.
The gradient of the $T$--$P$ coexistence curve corresponds to the change in the volume per particle $\upDelta v$ and the entropy per particle $\upDelta s$ via the Clapeyron equation, $\partial T/\partial P=\upDelta v / \upDelta s=T\upDelta v / \upDelta h$, where $\upDelta h$ is the enthalpy change per particle.
The smaller gradient for the \sevenP{}--\sevenPp{} transition compared to that of the \sevenPp{}--liquid and \sevenP{}--liquid transitions could arise either from a larger numerator or a smaller denominator, or both.
Using the data plotted in \figrefsub{fig:lowT-classification}{c--d}, we can compute, for example, that at \SI{1400}{\kelvin}, although $\upDelta v$ is about \SI{30}{\percent} larger for the \sevenPp{}--\sevenP{} transition that for the \sevenPp{}--liquid transition at their respective coexistence points, $\upDelta h$ is even larger, leading to an overall lower $\partial T/\partial P$.
It therefore appears that the enthalpy loss, and, in turn, the entropy gain of the highly superionic phase \sevenPp{} is the driving force for the \sevenP{}--\sevenPp{} phase transition, and is the reason that the melting point of the solid is considerably higher than it might otherwise be.
However, the liquid--solid coexistence line does not appear to have a significant change in gradient precisely at the triple point; indeed close to the triple point, all three phases have similar densities and enthalpies.

For the remaining ice phases (VII, \sevenP{} (T), \sevenP{} (R), X) shown in \figrefsub{fig:lowT-classification}{a}, we observe no clear discontinuities in density or enthalpy, in agreement with experiment~\cite{Hemley1987, Wolanin1997, Loubeyre1999} and previous simulation work~\cite{French2015,Trybel2020}.
Moreover, none of these phases appears to be metastable with respect to the others in this class. 
It is therefore not possible to determine the order of such phase transitions by computing chemical potentials, determining where they cross over and computing their gradients at the binodal point.
Instead, we compute the chemical potential by thermodynamic integration along isotherms and isobars across all phases of interest.
In principle, thermodynamic integration can only be performed along reversible paths, i.e.~without any first-order phrase transitions along the way~\cite{Vega2008}.
If we are able to integrate the chemical potential along a `closed loop' in $T$--$P$ space, we therefore expect that, should any first-order transitions occur along the way, the integral will diverge and we will obtain different values for the chemical potential for the same point when arrived at along different pathways.
We have computed the chemical potential along several such pathways, including that indicated by the example route in \figrefsub{fig:lowT-classification}{a}.
In all cases, we obtain the same chemical potential within numerical error for the same point even when integrated over large regions of $T$--$P$ space and over any combination of the phases shown in yellow, red and green in \figrefsub{fig:lowT-classification}{a}.
We show an example of the range of chemical potentials along the example TI route in  \figrefsub{fig:lowT-classification}{f}.
Within the numerical accuracy of free-energy calculations, it therefore appears that  ices VII, \sevenP{} (T), \sevenP{} (R) and X are not thermodynamically distinct, but merely different manifestations of the same thermodynamic phase under different conditions.

\section{X--{\protect\Pbcm{}} transition}

Below the superionic transition temperature, \Pbcm{} is believed, based on geometry optimization at zero temperature using DFT, to become stable over X at pressures higher than $\sim$\SI{350}{\giga\pascal}~\cite{Benoit1996,Caracas2008, Cheng2021}.
However, the precise location of the phase boundary at finite temperatures with the inclusion of nuclear quantum effects (NQEs) is unknown, and the mechanism of this X--\Pbcm{} transition is not clear.
Here we first provide a static-lattice picture of the ice phases at the DFT level,
and then exploit the MLP, thermodynamic integration and enhanced sampling to provide both thermodynamic and kinetic descriptions of the X--\Pbcm{} transition.

\paragraph{\SI{0}{\kelvin} enthalpy curves for ice phases at $P>\SI{100}{\giga\pascal}$}
First we present the \SI{0}{\kelvin} description for different ice phases predicted by the PBE DFT functional.
We considered all the ice structures reported in Ref.~\cite{Cheng2021}, which were identified by performing ab initio random structure searches (AIRSS)\cite{Pickard2011}.
We computed the \SI{0}{\kelvin} enthalpy at pressures from \SIrange{100}{1100}{\giga\pascal} for different ice phases predicted by the PBE DFT functional.
The MLP and revPBE0-D3 results are shown in \figrefSI{fig:enthalpy-curves-si}; they are qualitatively similar to the PBE results.

\begin{figure}
   \centering
     \includegraphics[width=0.45\textwidth]{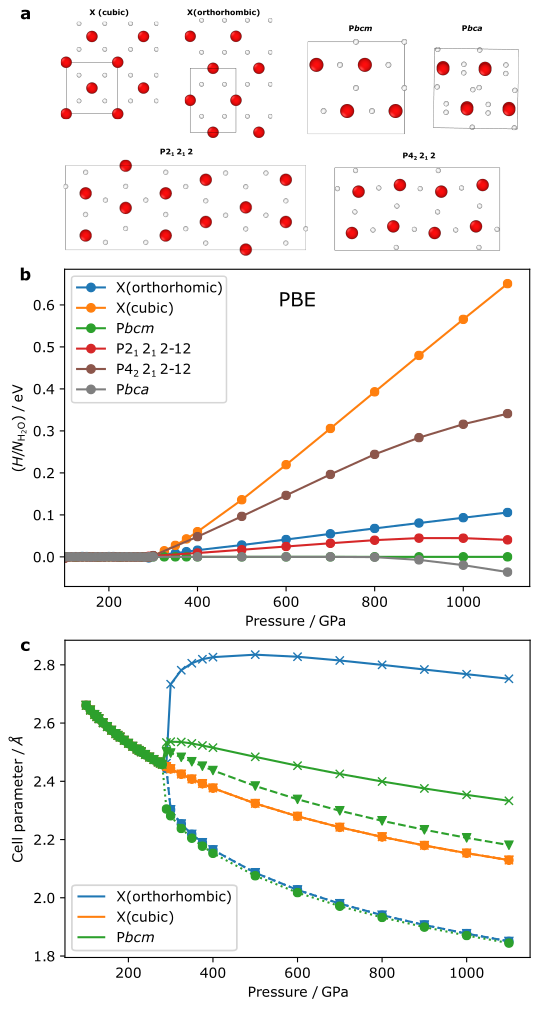}
   \caption{\textbf{Structures and properties of high-pressure phases at \SI{0}{\kelvin}.}
   \figLabelCapt{a} The structures of the ice phases considered.
   The simulation supercells used in DFT calculations are indicated using black boxes. 
   \figLabelCapt{b} \SI{0}{\kelvin} enthalpy curves for different ice phases predicted by the PBE DFT functional.
   \figLabelCapt{c} The length of the edge vectors of the simulation supercell ($l_{xx}$, $l_{yy}$ and $l_{zz}$) for X(orthorhombic), X(cubic) and \Pbcm{} phases at \SI{0}{\kelvin}.
   Notice that all three cell parameters for X(cubic) overlap,
   two edge lengths for the X(orthorhombic) cell are the same,
   and for \Pbcm{} there are three different cell parameters.
   The simulation supercells used for X(cubic) and X(orthorhombic) have 2 water molecules, while the \Pbcm{} supercell has 4 molecules (see panel \figLabelCapt{a}), so the cell parameters for the \Pbcm{} supercell in two directions have been divided by $\sqrt{2}$.}
   \label{fig:enthalpy-curves}
\end{figure}

In \figrefsub{fig:enthalpy-curves}{b} we show the PBE enthalpy curves for the 
P4$_2$/\textit{nnm} (ice X(orthorhombic)),  P\textit{n}$\overline{3}$\textit{m} (ice X(cubic)), \Pbca{}, \Pbcm{}, \PtwoOne{} and \PfourTwo{} phases, whose structures are illustrated in \figrefsub{fig:enthalpy-curves}{a}.
The \PtwoOne{} and \PfourTwo{} structures in \figrefsub{fig:enthalpy-curves}{b} can be considered to be mixed-stacking low-temperature insulating ice structures, and they have excess \SI{0}{\kelvin} enthalpies compared to the \Pbcm{} phase.
Both \Pbcm{} and \Pbca{} have a distorted hcp oxygen lattice and differ only in the position of hydrogen atoms at $P>\SI{800}{\giga\pascal}$.
In what follows, we focus just on the \Pbcm{} phase at $100<P/\si{\giga\pascal}<800$, as \Pbca{}, \PtwoOne{} and \PfourTwo{} are either the same or simply have a different stacking sequence in this pressure range.

X(orthorhombic) and X(cubic) both have a body-centred oxygen lattice; however, the cubicity of their unit cells differs above \SI{300}{\giga\pascal}.
Our phonon calculations [\figrefSI{fig:SI-phonon-DOS}] further show that the X(cubic) structure is at a saddle point of the potential-energy surface of the system above $\sim$\SI{400}{\giga\pascal}, as suggested by its imaginary vibrational modes.
Such a dynamic instability in cubic ice X was previously reported in Ref.~\cite{Caracas2008}.
Given that the orthorhombic X structure also has a considerably lower enthalpy, we therefore regard it to be the \emph{de facto} X phase at above \SI{300}{\giga\pascal}.
In other words, below \SI{300}{\giga\pascal},
ice X has a fully cubic cell (space group P\textit{n}$\overline{3}$\textit{m}), and above
\SI{300}{\giga\pascal} the cell shape is orthorhombic (space group P4$_2$/\textit{nnm}).
We use `X' to refer to the phase with the cubic--orthorhombic transition at \SI{300}{\giga\pascal}.

In the pressure range from \SIrange{100}{300}{\giga\pascal}, all the phases considered collapse to the cubic ice-X structure, while the enthalpy difference gradually increases at higher pressures.
To probe this collapse further, we show the change in lattice constants as a function of pressure for the X(cubic), X(orthorhombic) and \Pbcm{} structures in \figrefsub{fig:enthalpy-curves}{c}.
At $P<\SI{300}{\giga\pascal}$, the cell parameters of the \Pbcm{} and the X(orthorhombic) structure indeed correspond to those of X(cubic).
At above \SI{300}{\giga\pascal}, the shape of the cell, which is related to the configuration of the ice structures, exhibits a sharp, first-order-like change at the transition.
By contrast, the enthalpy (\figrefsub{fig:enthalpy-curves}{b}), the potential energy and the molar volume (\figrefSI{fig:SI-zoomed-curves}) all exhibit rather smooth changes across the onset of the bifurcation at \SI{300}{\giga\pascal}.
Moreover, although the configurational change is sharp, the transition happens readily during geometry optimization at \SI{0}{\kelvin}, suggesting a lack of an activation barrier associated with the transitions both between X(orthorhombic) and X(cubic) and between \Pbcm{} and X(cubic) at the bifurcation point.
To the best of our knowledge, such a bifurcation point has not been observed before in any other systems.

\begin{figure*}
   \centering
\includegraphics[width=0.75\textwidth]{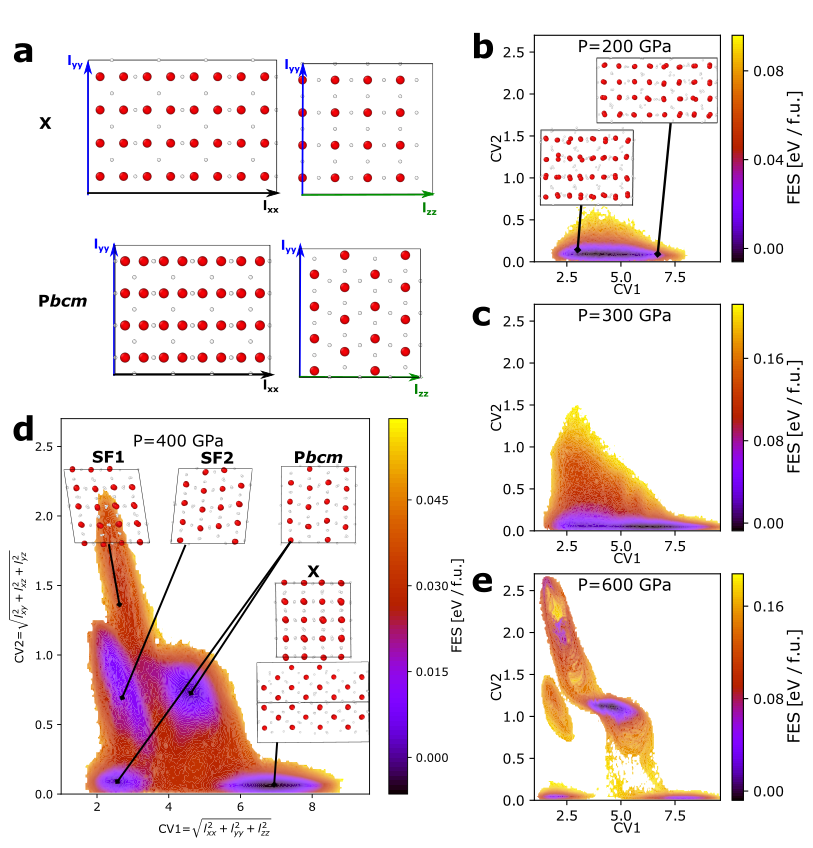}
   \caption{\textbf{Free energy of the X--\Pbcm{} transition.}
   \figLabelCapt{a} Configurations of \Pbcm{} and X.
      The orthorhombic simulation cell contains 64 water molecules and is commensurate with both the X and the \Pbcm{} structures.
   Free-energy surfaces (FES) at \figLabelCapt{b} \SI{200}{\giga\pascal}, \figLabelCapt{c} \SI{300}{\giga\pascal}, \figLabelCapt{d} \SI{400}{\giga\pascal} and \figLabelCapt{e} \SI{600}{\giga\pascal}, all computed at \SI{1000}{\kelvin}, as a function of two collective variables: CV$_1$ measures the anistropy of the box and CV$_2$ measures the tilt of the box.
   }
   \label{fig:metad}
\end{figure*}

\paragraph{X--\Pbcm{} transition at finite temperatures}
To understand better the nature of the X--\Pbcm{} phase transition, we investigated its behaviour at finite temperatures.
To this end, we ran well-tempered metadynamics simulations with adaptive bias employing the MLP in the $NPT$ ensemble at a temperature of \SI{1000}{\kelvin}, with all the simulation supercell parameters allowed to fluctuate.
We employed an orthorhombic supercell of 64 molecules, illustrated in \figrefsub{fig:metad}{a}.
The shape and size were selected such that the supercell was commensurate with both the X and the \Pbcm{} ice structures,
as well as other \Pbcm{}-like structures with mixed stacking.
We used two collective variables (CVs):
$\text{CV}_1 = [l_{xx}^2+l_{yy}^2+l_{zz}^2]^{1/2}$ is based on the anisotropy of the edge vectors of the simulation supercell, whilst
$\text{CV}_2  = [l_{xy}^2+l_{xz}^2+l_{yz}^2]^{1/2}$ measures the tilt of the supercell, whose edge vectors are defined as $(l_{xx},\,l_{yx},\,l_{zx})$, $(l_{xy},\,l_{yy},\,l_{zy})$ and $(l_{xz},\,l_{yz},\,l_{zz})$.
It is worth noting that, even if the geometry of the supercell is carefully selected, when probing solid--solid transitions using metadynamics, there can be some hysteresis which affects the computed free-energy profile.

The computed free-energy surfaces (FES) at \SI{200}{\giga\pascal}, \SI{300}{\giga\pascal}, \SI{400}{\giga\pascal} and \SI{600}{\giga\pascal} and a temperature of \SI{1000}{\kelvin} are presented in \figref{fig:metad}.
At $P=\SI{400}{\giga\pascal}$, the FES has five distinct minima:
one for ice X, two for \Pbcm{}, and two corresponding to structures similar to \Pbcm{} but with different stacking sequences.
During the metadynamics run, the system had frequent transitions between the five states.
The regions that can be considered as transition states are wide and only have a free-energy excess of $\sim$\SI{0.03}{\electronvolt} per formula unit (f.u.).
At $P=\SI{600}{\giga\pascal}$, the FES still has the five minima that correspond to the four distinct ice structures,
although the minima are more separated and the transition states have much higher free energy ($\sim$\SI{0.18}{\electronvolt} per f.u.).
At $P=\SI{300}{\giga\pascal}$, the regions associated with the five minima are instead connected with no transition barrier.
At $P=\SI{200}{\giga\pascal}$, only one minimum is explored in our metadynamics simulations even when a substantial amount of a biasing potential has been added.
This minimum corresponds to the X phase, and the main variance within the minimum is associated with the elongation of the supercell.

The FES in \figref{fig:metad} again confirms the bifurcation transition revealed in the \SI{0}{\kelvin} enthalpy curves in \figref{fig:enthalpy-curves}:
at high pressures ($P \ge \SI{300}{\giga\pascal}$), there are distinct phases including \Pbcm{}, X and the phases with stacking faults, while at low pressures ($P \le \SI{300}{\giga\pascal}$), all the phases collapse into X.
These solid--solid transitions are more facile at lower pressures, as suggested by the lack of activation barriers at \SI{300}{\giga\pascal} and the lower barriers at \SI{400}{\giga\pascal} compared to \SI{600}{\giga\pascal}.
The transition pathway from X to \Pbcm{} involves a highly collective shuffling of the \{100\} planes in X, similar to the bcc--hcp transitions in many metals~\cite{Benoit1996, Militzer2010}.

\paragraph{Finite-temperature phase diagram}

Both the static lattice enthalpy curves and the results from finite-temperature metadynamics simulations suggest that \Pbcm{} and X are distinct phases above $\sim$\SI{300}{\giga\pascal}, but collapse into a single phase at lower pressures.
Where they are two separate phases, one can ask what their relative thermodynamic stabilities are and how the phase boundary depends on temperature.
Moreover, the \SI{0}{\kelvin} enthalpy curves only provide a partial description of the chemical potentials: although \Pbcm{} has a lower enthalpy at the static-lattice level, the excess enthalpy of the X phase is rather small, and thermal and quantum fluctuations can have a strong influence on the thermodynamic free energies for water systems~\cite{Cheng2019, Reinhardt2021}.

Although X and \Pbcm{} appear to be identical below \SI{300}{\giga\pascal}, during the free-energy calculations using thermodynamic integration, we choose to be agnostic about this fact and instead always treat \Pbcm{} and X separately at all pressures.
Generally speaking, when computing the classical Gibbs energy of a system, TI can be performed between the reference crystal and the classical physical system at a certain $P$-$T$ point (the so-called $\lambda$-TI, or hamiltonian TI~\cite{Frenkel1984}),
or along isotherms and isobars to calculate the chemical-potential differences at other pressure and temperature conditions.
It is worth noting that, as the X(cubic)--X(orthorhombic) and X(cubic)--\Pbcm{} transitions at about $\sim$\SI{300}{\giga\pascal} are peculiar and accompanied by sharp jumps in the lattice parameter (\figrefsub{fig:enthalpy-curves}{c}), although the enthalpy changes are smooth, it is not completely clear whether the TI method still applies along the pressure across the transition boundary.
To clarify this ambiguity, we performed multiple TI calculations using the MLP starting from the reference crystals at different pressures from \SIrange{200}{700}{\giga\pascal} in \SI{100}{\giga\pascal} steps, and at several temperatures (\SI{100}{\kelvin}, \SI{300}{\kelvin}, \SI{600}{\kelvin}, \SI{1000}{\kelvin}).
As shown in \figrefSI{fig:SI-mu-x-pbcm},
these different TI routes provide consistent estimations of the classical chemical potentials for both phases employing the MLP,
which not only verifies that these values are statistically converged,
but also suggests that TI can be used to cross the transition boundary at $\sim$\SI{300}{\giga\pascal}.
This again highlights the peculiarity of the bifurcation transition, which has continuous enthalpy and free energy, but sharp changes in structures.

\begin{figure}
   \centering
   \includegraphics[width=0.45\textwidth]{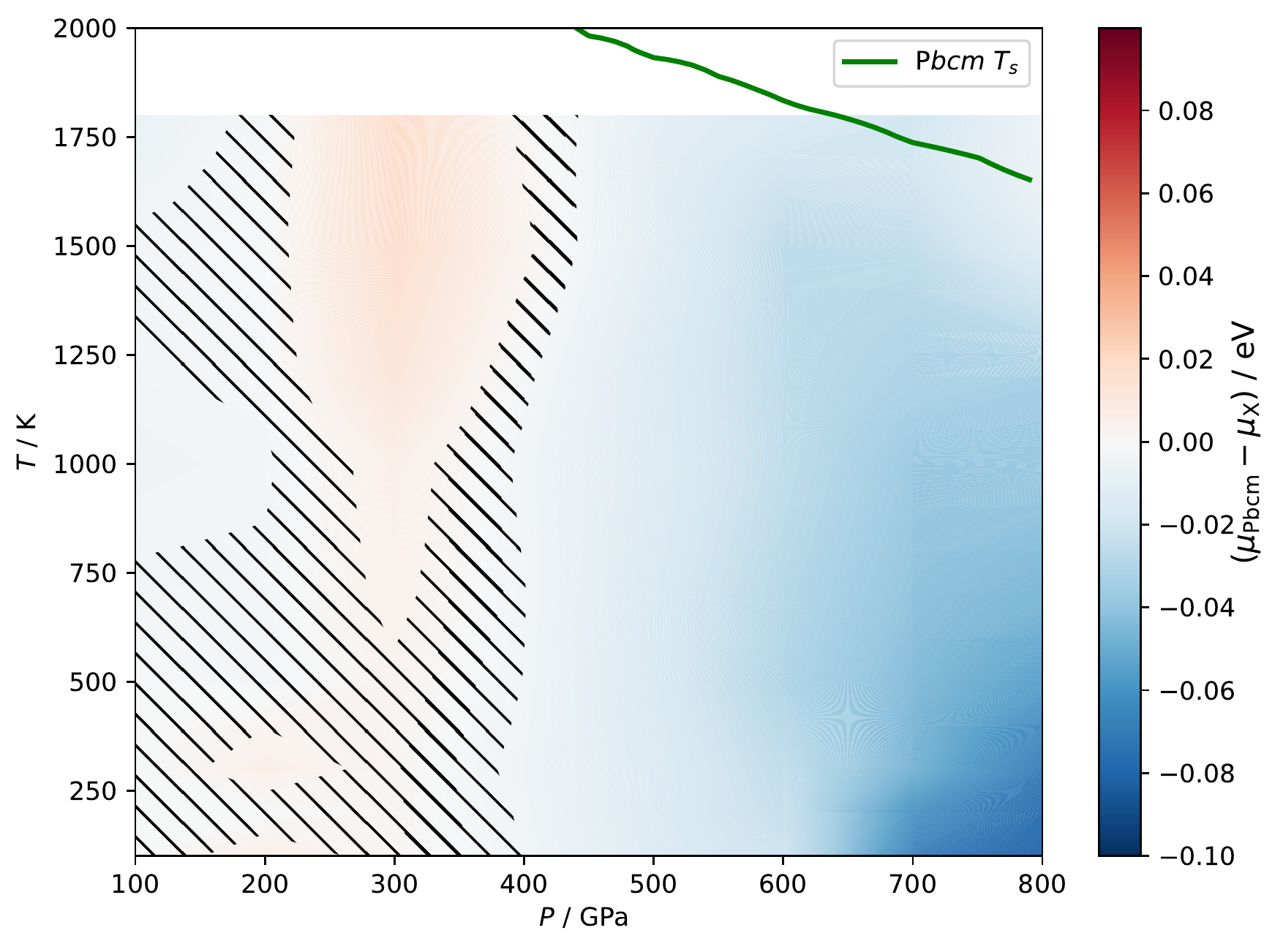}
   \caption{
   \textbf{High-pressure phase diagram.}
   The chemical-potential difference $\mu_\text{\Pbcm{}}-\mu_\text{X}$ per formula unit (f.u.)  between \Pbcm{} and X as a function of temperature and pressure.
   The statistical error in the chemical-potential difference results in the uncertainty of the coexistence line, which is indicated by the hatched area.
   The green curve shows the insulating--superionic transition temperature $T_\text{s}$ defined by a threshold of proton diffusivity ($D_\text{H}>\SI{e-8}{\metre\squared\per\second}$) for the \Pbcm{} phase.
   }
   \label{fig:X-Pbcm-phasediagram}
\end{figure}

To determine the chemical potentials of X and \Pbcm{} at the PBE level that include NQEs, we add together the classical chemical potential at the MLP level, a correction term in the free energies that removes the small residual errors in the MLP~\cite{Behler2015,Cheng2019}, as well as a term that accounts for the influence of NQEs.
The resulting phase diagram is shown in \figref{fig:X-Pbcm-phasediagram}.
The chemical-potential difference between the two phases is generally very small across the whole temperature and pressure range that we consider.
There is essentially no difference between the chemical potentials of X and \Pbcm{} at $P<\SI{400}{\giga\pascal}$.
Indeed, the structures of X and \Pbcm{} at low pressures are the same in that region.
The small difference at higher temperatures and low pressures is due to the numerical error in the integration. 
However, at higher pressures, \Pbcm{} becomes more stable.
Within statistical uncertainty, the finite-temperature phase diagram is thus consistent with the \SI{0}{\kelvin} enthalpy results of \figrefsub{fig:enthalpy-curves}{a}.  
There is a weak temperature dependence of the chemical potentials, with higher $T$ slightly favouring the stability of X.

\section{Discussion}
In this study, we combined a PBE DFT description and a machine-learning potential to investigate the nature of the phase transitions between high-pressure phases of insulating ice.
Crucially, our approach probes the thermodynamic behaviours of the ice phases.
By accurately computing the chemical-potential differences between the phases, we can classify whether these phases are thermodyanmically distinct, and determine the orders of the phase transitions between them.

The superionic transition in bcc ice phases has been studied in both theoretical and experimental work~\cite{Cavazzoni1999,Hernandez2016, Hernandez2018, Queyroux2020, Cheng2021, Sugimura2012,Millot2018, Lin2005, Schwager2004,Schwager2008}, although the range of conditions at which superionic phases are observed is subject to considerable variation in the literature.
Although there is a degree of superionicity in ice \sevenP{}, the hydrogen diffusivity increases significantly in ice \sevenPp{}~\cite{Hernandez2018}.
In experiment, several signatures of a first-order-like phase transition have been reported, particularly in the context of a discontinuity in the gradient of the $P$--$T$ coexistence curve between the liquid and a solid phase, which has sometimes been interpreted as a triple point.
This has been reported under a range of different conditions, from \SI{35}{\giga\pascal} and \SI{1040}{\kelvin}~\cite{Lin2005} to \SI{43}{\giga\pascal} and $\sim$\SI{1600}{\kelvin}~\cite{Schwager2004} to \SI{14.6}{\giga\pascal} and \SI{850}{\kelvin}~\cite{Queyroux2020}, and the solid phases that apparently coexist with the liquid have occasionally been reported to be different, e.g.~VII and X in Ref.~\citenum{Schwager2004}.
We have obtained a triple point between the liquid, \sevenP{} and \sevenPp{} at $\sim$\SI{20}{\giga\pascal} and \SI{875}{\kelvin}, close to the value reported by Queyroux~\textit{et al.}~\cite{Queyroux2020}; moreover, we have shown that \sevenP{} and \sevenPp{} are thermodynamically distinct phases, as suggested by the apparent (though often small) discontinuities of the enthalpy and density in simulations using system sizes large enough to investigate thermodynamic behaviour.
We have explicitly computed their chemical potentials to confirm that the \sevenP{}--\sevenPp{} phase transition is of first order, and we determined the associated thermodynamic coexistence curve between them.
We speculate that one possible reason for the various different reported triple points is that in our simulations, it does not appear that the solid--liquid coexistence curve changes gradient significantly precisely at the triple point; there is a more gradual change of gradient as higher pressures and temperatures are reached, as the supercritical ice phase becomes progressively more different from its low-temperature less-disordered analogue.

In X-ray diffraction experiments, no clear distinction has been seen between the other bcc ice phases, including VII and X~\cite{Hemley1987, Wolanin1997, Loubeyre1999}.
We confirm here that ices VII, \sevenP{} (T), \sevenP{} (R) and X show no clear discontinuities in thermodynamic quantities including density and enthalpy, and in turn the chemical potentials.
The only differences between them come from the positions and the dynamics of the hydrogen atoms.
We thus regard these phases to be merely different manifestations of the same thermodynamic phase under different conditions.

For the insulating ice phases that have been theoretically predicted for the pressure range between \SI{100}{\giga\pascal} and \SI{800}{\giga\pascal}, we find that all of them (X(orthorhombic), X(cubic), \Pbca{}, \Pbcm{}, \PtwoOne{} and \PfourTwo{}) collapse into ice X below \SI{300}{\giga\pascal}.
This means that there is no distinction between them at pressures from \SIrange{100}{300}{\giga\pascal}.
Moreover, the transition barriers between X, \Pbcm{} and the structures with stacking faults increase as the pressure increases.
Above \SI{300}{\giga\pascal}, X and \Pbcm{} are thermodynamically distinct phases, and we have computed their relative chemical potentials, fully accounting for thermal and nuclear fluctuations.
\Pbcm{} has not yet been experimentally observed, despite past and on-going dynamical compression experiments.
Our computed phase diagram in ~\figref{fig:X-Pbcm-phasediagram} along with the peculiar solid--solid mechanism we have identified provide the following predictions to guide future experiments:
(i) 
The phase boundary between \Pbcm{} and X is largely independent of temperature.
By contrast, as suggested in Ref.~\cite{Cheng2021}, the \Pbcm{} phase becomes superionic and 
its distorted hexagonal close-packed (hcp) oxygen lattice becomes fully hcp
at the superionic transition temperature $T_\text{s}$.
This $T_\text{s}$, which is marked by the green curve in \figref{fig:X-Pbcm-phasediagram}, is $\sim$\SI{1800}{\kelvin} between \SI{300}{\giga\pascal} and \SI{800}{\giga\pascal}.
Above this temperature, the superionic fcc phase (ice~XVIII~\cite{Millot2019}) and the superionic hcp phase were predicted to have similar chemical potentials and were both regarded as thermodynamically stable between about \SI{100}{\giga\pascal} to \SI{800}{\giga\pascal}~\cite{Cheng2021}.
This means, if feasible, keeping the temperature low (at least below the superionic transition temperature) is essential for making the \Pbcm{} phase in compression experiments.
(ii) Between \SI{300}{\giga\pascal} and \SI{800}{\giga\pascal}, 
the chemical-potential difference $\mu_\text{X}-\mu_\text{\Pbcm{}}$ increases with increasing pressure, although the magnitude of the chemical-potential difference remains rather small even at \SI{800}{\giga\pascal}.
This means the equilibrium driving force favouring the thermodynamic stability of the \Pbcm{} phase is greater at higher pressures.
On the other hand, the activation barrier for the X--\Pbcm{} transition also becomes larger, meaning that the kinetic transition rate is smaller at higher pressures.
This implies that, in compression experiments that aim to make the \Pbcm{} phase, instead of
targeting the highest pressure achievable, it may be advantageous to aim for pressures that are moderately above the X--\Pbcm{} transition line in order to utilize the lower kinetic barrier.
(iii) If the \Pbcm{} phase, or a mixed-stacking phase, is indeed made during compression experiments, it will not remain stable and will instead reversibly transform into X whenever the external pressure drops to below \SI{300}{\giga\pascal}.

\section{Conclusions}

In summary, our study provides a thermodynamic view of the high-pressure phase diagram of ice below the superionic transition line.
Our results demonstrate that the various bcc ice phases, including ice VII, ice \sevenP{} (T/R) and ice X, are indeed the same thermodynamic phase that behaves differently at different conditions.
Our simulations also reveal the peculiar bifurcation and transition pathway between X and \Pbcm{}, which sheds light on the viability of the experimental synthesis of the as yet experimentally undetected, but long predicted, \Pbcm{} phase.
Moreover, our methodological framework, which combines first-principles methods, machine-learning potentials, free-energy methods and enhanced sampling, can be applied to study the phase diagrams and phase transitions of other polymorphic systems, such as high-pressure solid hydrogen~\cite{Cheng2020}, perovskites~\cite{Jinnouchi2019} and two-dimensional materials~\cite{Miro2014}.

\section{Methods}

\paragraph{DFT calculations}
The DFT calculations were carried out with VASP 5.4 using the hard PAW pseudopotentials and an energy cutoff of \SI{1000}{\electronvolt}. The $k$ points were chosen automatically using a $\upGamma$-point-centred grid with a resolution of \SI{0.3}{\per\angstrom}.
We considered nine structures with the space groups P4$_2$/\textit{nnm} (ice X(orthorhombic)), P3$_2$21, P4$_3$22, P2$_1$2$_1$2, P4$_2$2$_1$2, \Pbca{}, \Pbcm{} and P\textit{n}$\overline{3}$\textit{m} (ice X(cubic)), which were taken from a previous study~\cite{Cheng2021}. 
That study predicted the structures with AIRSS~\cite{Pickard2011} using CASTEP to perform the DFT calculations and applying various exchange-correlation functionals.
Their identified PBE structures were reoptimized in this study with VASP at constant pressure to remove any residual forces.
We considered a pressure range between \SI{100}{\giga\pascal} and \SI{1100}{\giga\pascal} and checked that the initial space group for each considered structure was maintained after reoptimization.
The energies were typically converged to \SI{e-8}{\electronvolt} and the forces to \SI{e-6}{\electronvolt\per\angstrom}.
Subsequently, the structures were optimized similarly with the revPBE0-D3 exchange-correlation functional to investigate the influence of a hybrid functional on the structural stability.
Finally, we calculated the phonons for the P4$_2$/\textit{nnm}, \Pbcm{} and P\textit{n}$\overline{3}$\textit{m} structures at the PBE level to identify possible pressure ranges of dynamical instability.
Those calculations were performed with Phonopy~\cite{Togo2015} on \num{2x2x2} supercells in \SI{25}{\giga\pascal} steps between \SI{100}{\giga\pascal} and \SI{400}{\giga\pascal}.
The phonon densities of states were evaluated with at least 71 mesh $q$ points in each direction.

\paragraph{MLP MD simulation details}

We performed $NPT$ MD simulations for bcc-ice systems with LAMMPS~\cite{Plimpton1995} with an MLP implementation~\cite{Singraber2019}.
Bulk phases were simulated in the $NPT$ ensemble using the Nos\'{e}--Hoover isotropic barostat~\cite{Shinoda2004} and a timestep of \SI{0.5}{\femto\second}, with typical MD simulation runs of \SI{50}{\pico\second} at each set of conditions considered.
Bulk-simulation system sizes ranged between 128 and 2048 water molecules to check for finite-size effects; for direct-coexistence simulations, we used system sizes of between 2916 and 3453 molecules.

To compute chemical potentials and in turn the phase diagram, we started by considering a large direct-coexistence simulation~\cite{Opitz1974} of liquid water in contact with ice and determined the coexistence temperature at pressures of \SI{20}{\giga\pascal} and \SI{26}{\giga\pascal}, and the coexistence pressure at \SI{1800}{\kelvin}.
These points match very well the coexistence curve reported in Ref.~\citenum{Cheng2021}.
At these coexistence points, the liquid and the appropriate solid phases have the same chemical potentials.
Finally, we computed thermal averages of system properties along isotherms and isobars in MD simulations, and numerically integrated~\cite{Vega2008} the Gibbs--Duhem relation at constant temperature to give the chemical potential as a function of pressure $\mu(P) = \mu(P_1) + \int_{P_1}^P v(P')\,\mathrm{d} P'$,
where $v$ is the volume per formula unit,  and the Gibbs--Helmholtz equation at constant pressure to give the chemical potential as a function of temperature,
$\beta \mu(T) = \beta_1 \mu(T_1) - \int_{T_1}^{T} (H/Nk_\text{B}(T')^2)\,\mathrm{d} T'$,
where $\beta=1/k_\text{B}T$ and $H$ is the enthalpy.
This enthalpy includes kinetic energy contributions, i.e.~the de Broglie thermal wavelength's temperature dependence is accounted for in the chemical potential~\cite{Reinhardt2019}.
In each case, we fitted the integrands to a polynomial to be able to extrapolate them slightly beyond the regions of metastability.
We confirmed that starting from any of the direct-coexistence starting points, we can recover the other two, and others, by this thermodynamic integration route.
To find other coexistence points, we computed the chemical potentials for all phases that are metastable at a given temperature and pressure, and determined numerically the point at which the chemical potentials of different phases cross over.

\paragraph{The chemical potentials of X and \Pbcm{} phases}

We first computed the classical Gibbs energy for both \Pbcm{} and X using a system size of 512 formula units.
We performed geometry optimization for each structure at different pressures. 
Where a local minimum is found, the hessian matrices of the structures are computed using the program i-PI~\cite{Kapil2019} with a finite-difference method.  
The hessian matrix enables us to construct a reference harmonic crystal in which the forces between the atoms are defined by the phonon modes.
After that, we ran hamiltonian TI between the reference crystal and the classical physical system.
We used TI along isotherms and isobars to calculate the chemical-potential differences at other pressure and temperature conditions.
To validate that we can reach consistent estimates of the chemical potential using different TI pathways, we therefore performed multiple TI calculations:
(i) $\lambda$-TI at different temperatures (\SI{100}{\kelvin}, \SI{300}{\kelvin}, \SI{600}{\kelvin}, \SI{1000}{\kelvin}) at pressure $P_0$ (\SI{200}{\giga\pascal}, \SI{300}{\giga\pascal}, \SI{400}{\giga\pascal}, \SI{500}{\giga\pascal}, \SI{600}{\giga\pascal}, \SI{700}{\giga\pascal}), 
(ii) integrate along $T$ between \SI{100}{\kelvin} and \SI{2000}{\kelvin},
and finally (iii) integrate along pressure between \SI{100}{\giga\pascal} and \SI{800}{\giga\pascal}.
As shown in \figrefSI{fig:SI-mu-x-pbcm}, these different TI routes provide consistent estimations of the classical chemical potentials based on the MLP.

In addition, we promoted the MLP results to the PBE level by adding $\mu-\mu^\mathrm{MLP}$ computed using the free-energy perturbation method, which removes the small residual errors in the MLP partly due to its lack of long-range electrostatics~\cite{Behler2015,Cheng2019}.
To account for the influence of NQEs on the free energies,
we ran path-integral molecular dynamics (PIMD) simulations on small systems with 64 \ce{H2O} formula units using LAMMPS in conjunction with the i-PI code~\cite{Kapil2019}.
These PIMD simulations were performed employing the MLP, with 32 beads for all phases considered at the relevant constant pressure and temperature conditions.
NQEs on the chemical potentials were taken into account by integrating the quantum centroid virial kinetic energy $\avg{E_\text{k}}$ with respect to the fictitious `atomic' mass $\ficm$ from the classical (i.e.~infinite) mass to the physical masses $m$~\cite{Ceriotti2013,Cheng2016nuclear,Cheng2018hydrogen}.
The contribution from NQEs to the chemical-potential difference is small ($<\SI{5}{\milli\electronvolt}$ per molecule).
These two contributions to the chemical-potential difference are illustrated in \figrefSI{fig:SI-mu-x-pbcm-all}.

\textbf{Acknowledgements}
We thank Chris Pickard for providing the initial structures of high-pressure ice phases and for useful advice.
AR and BC acknowledge resources provided by the Cambridge Tier-2 system operated by the University of Cambridge Research Computing Service funded by EPSRC Tier-2 capital grant EP/P020259/1.
MB was supported by the European Union within the Marie Sk\l{}odowska-Curie actions (xICE grant 894725) and acknowledges computational resources at North-German Supercomputing Alliance (HLRN) facilities.
SH and MM acknowledge support from LDRD 19-ERD-031 and computing support from the Lawrence Livermore National Laboratory (LLNL) Institutional Computing Grand Challenge program. FC acknowledges support from the US DOE Office of Science, Office of Fusion Energy Sciences.
Lawrence Livermore National Laboratory is operated by Lawrence Livermore National Security, LLC, for the U.S.~Department of Energy, National Nuclear Security Administration under Contract DE-AC52-07NA27344.

\textbf{Data availability statement}
All original data generated for the study are in the
SI repository (url to be inserted upon acceptance of the paper).

\clearpage

\renewcommand{\thepage}{S\arabic{page}}
\renewcommand{\thefigure}{S\arabic{figure}}
\renewcommand{\theequation}{S\arabic{equation}}
\makeatletter  
\renewcommand*{\thesection}{S\arabic{section}}
\renewcommand*{\thesubsection}{\thesection.\arabic{subsection}}
\renewcommand*{\p@subsection}{}
\renewcommand*{\thesubsubsection}{\thesubsection.\arabic{subsubsection}}
\renewcommand*{\p@subsubsection}{}
\renewcommand\@seccntformat[1]{\csname the#1\endcsname\quad} %
\makeatother

\raggedbottom

\section*{Supplementary information}

\begin{figure}[H]
 \centering
 \begin{tabular}{@{}p{0.45\textwidth}@{\quad}p{0.45\textwidth}@{}}
   \subfigimg[width=\linewidth]{\figLabel{a}}{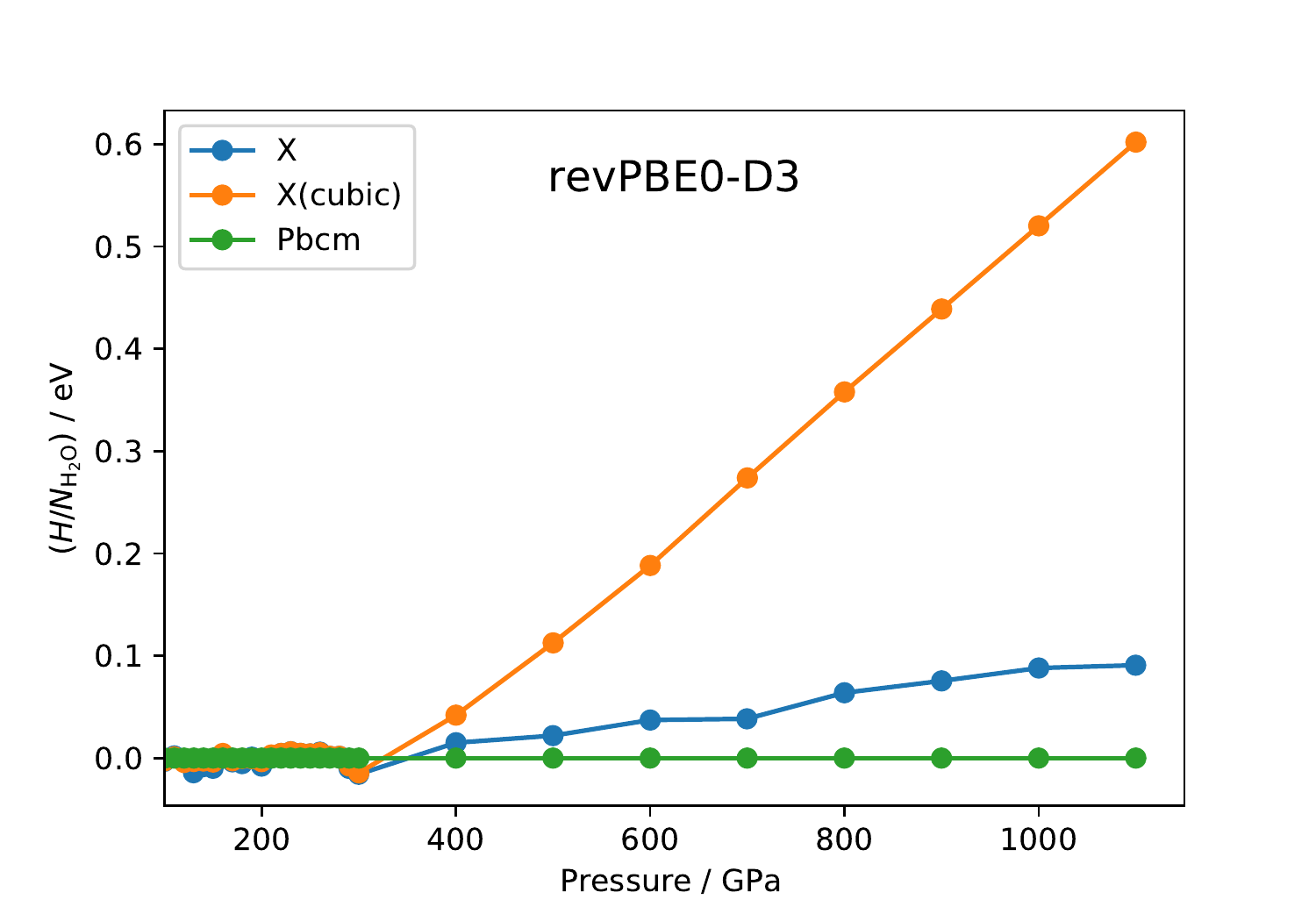} \\
   \subfigimg[width=\linewidth]{\figLabel{b}}{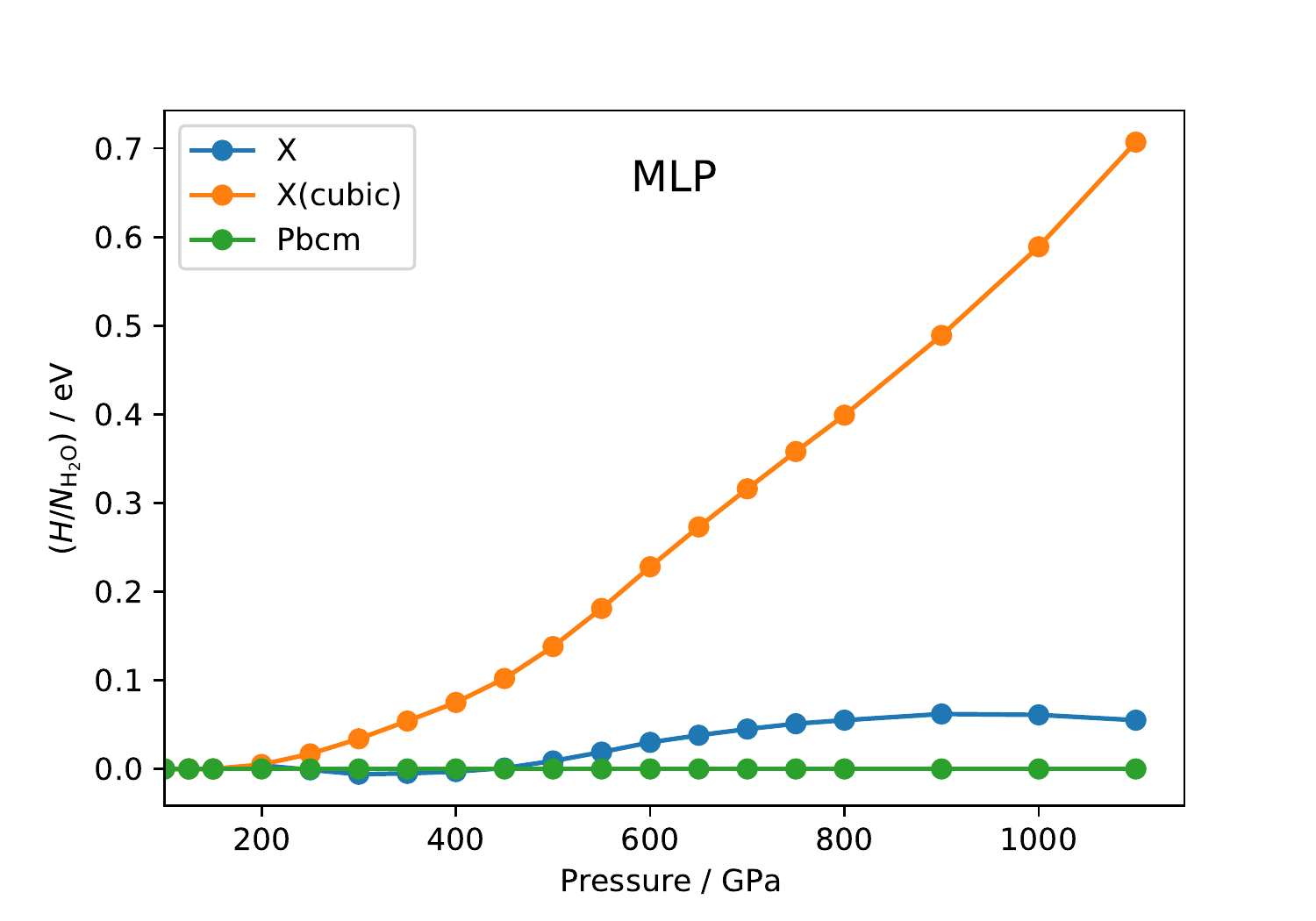}
 \end{tabular}
   \caption{\SI{0}{\kelvin} enthalpy curves for different ice phases predicted by \figLabelCapt{a} the revPBE0-D3 DFT functional and \figLabelCapt{b} the MLP.}
   \label{fig:enthalpy-curves-si}
\end{figure}

\begin{figure}[H]
   \centering
     \begin{tabular}{@{}p{0.45\textwidth}@{\quad}p{0.45\textwidth}@{}}
   \subfigimg[width=\linewidth]{\figLabel{a}}{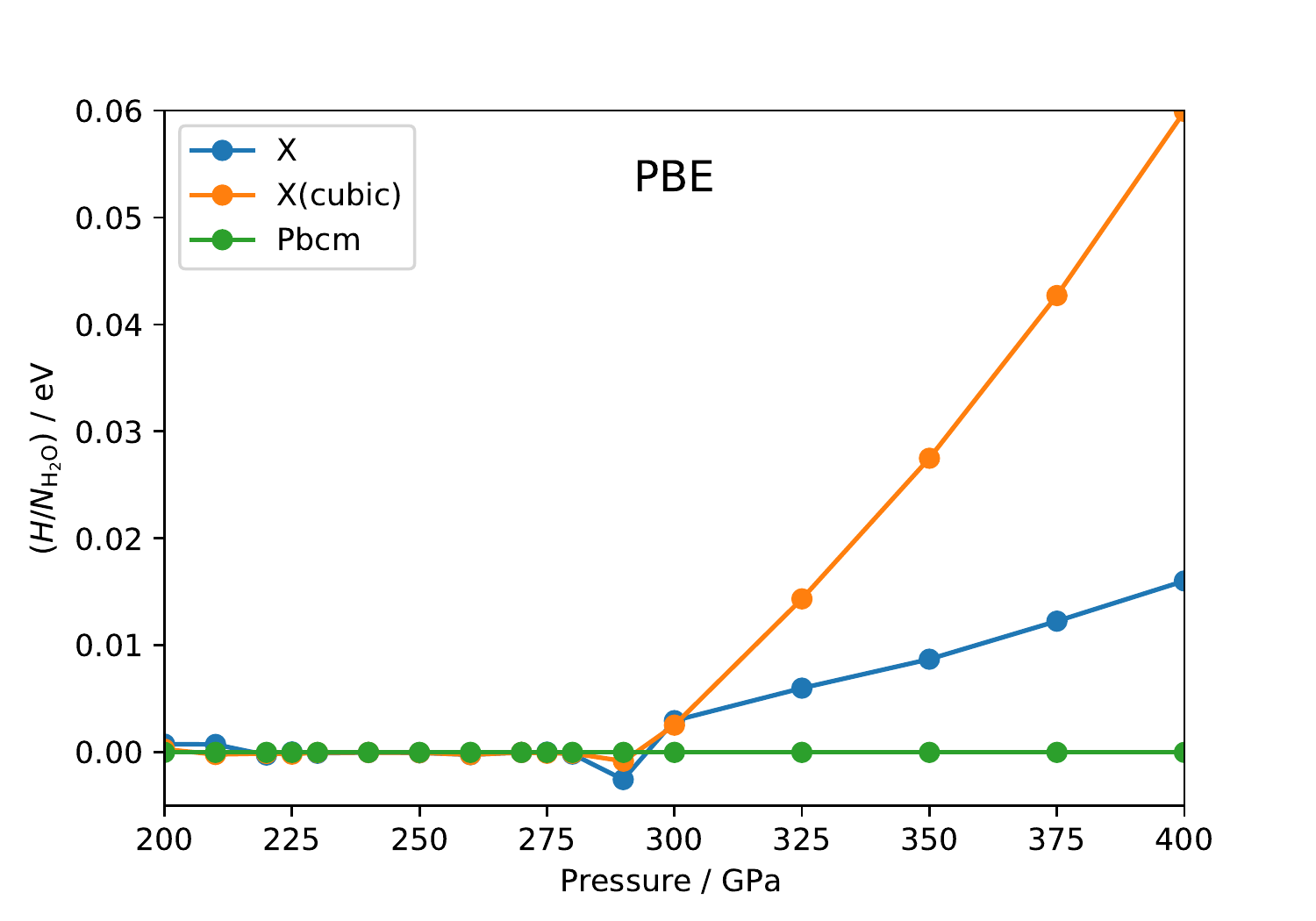} \\
   \subfigimg[width=\linewidth]{\figLabel{b}}{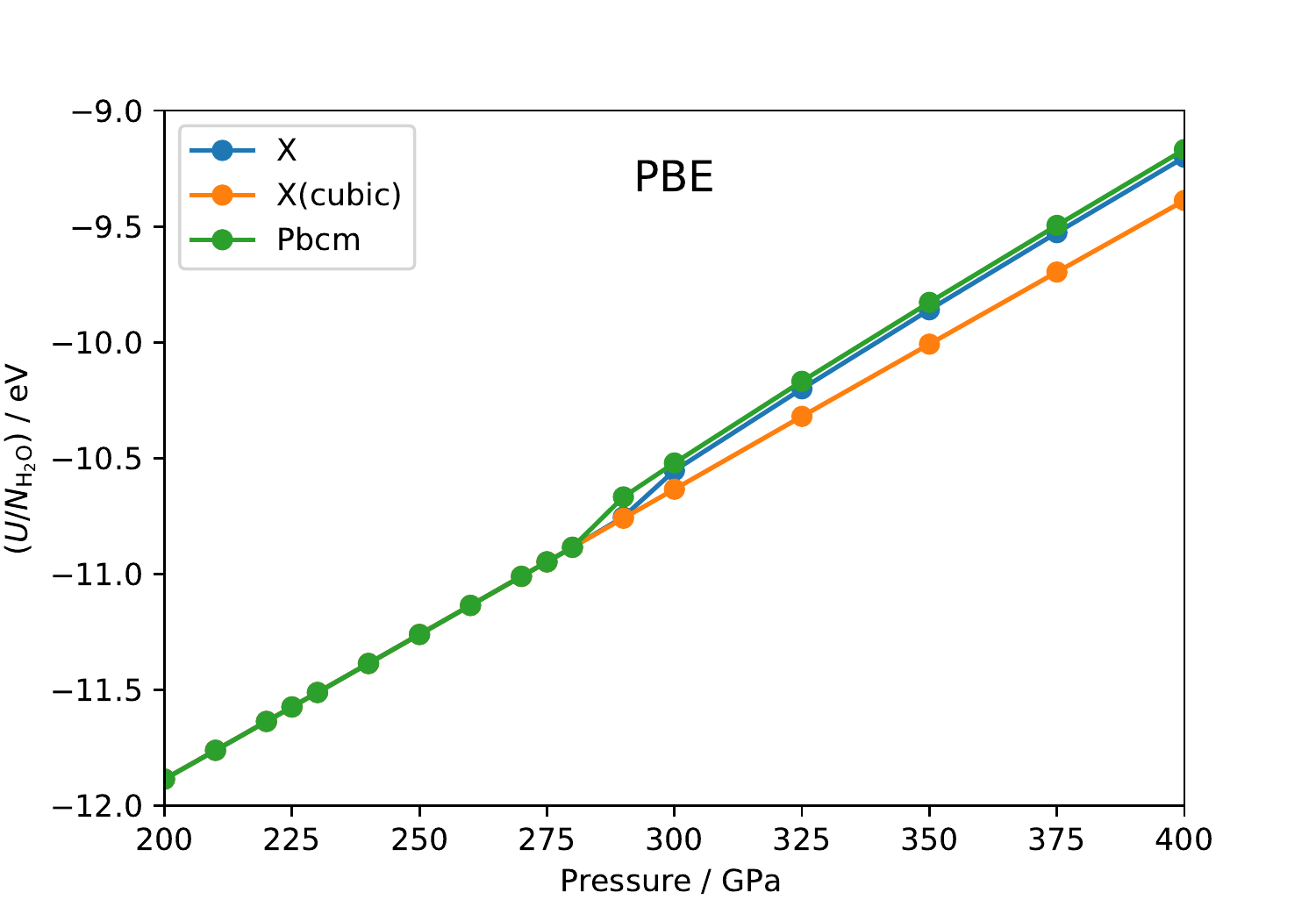}  \\
   \subfigimg[width=\linewidth]{\figLabel{c}}{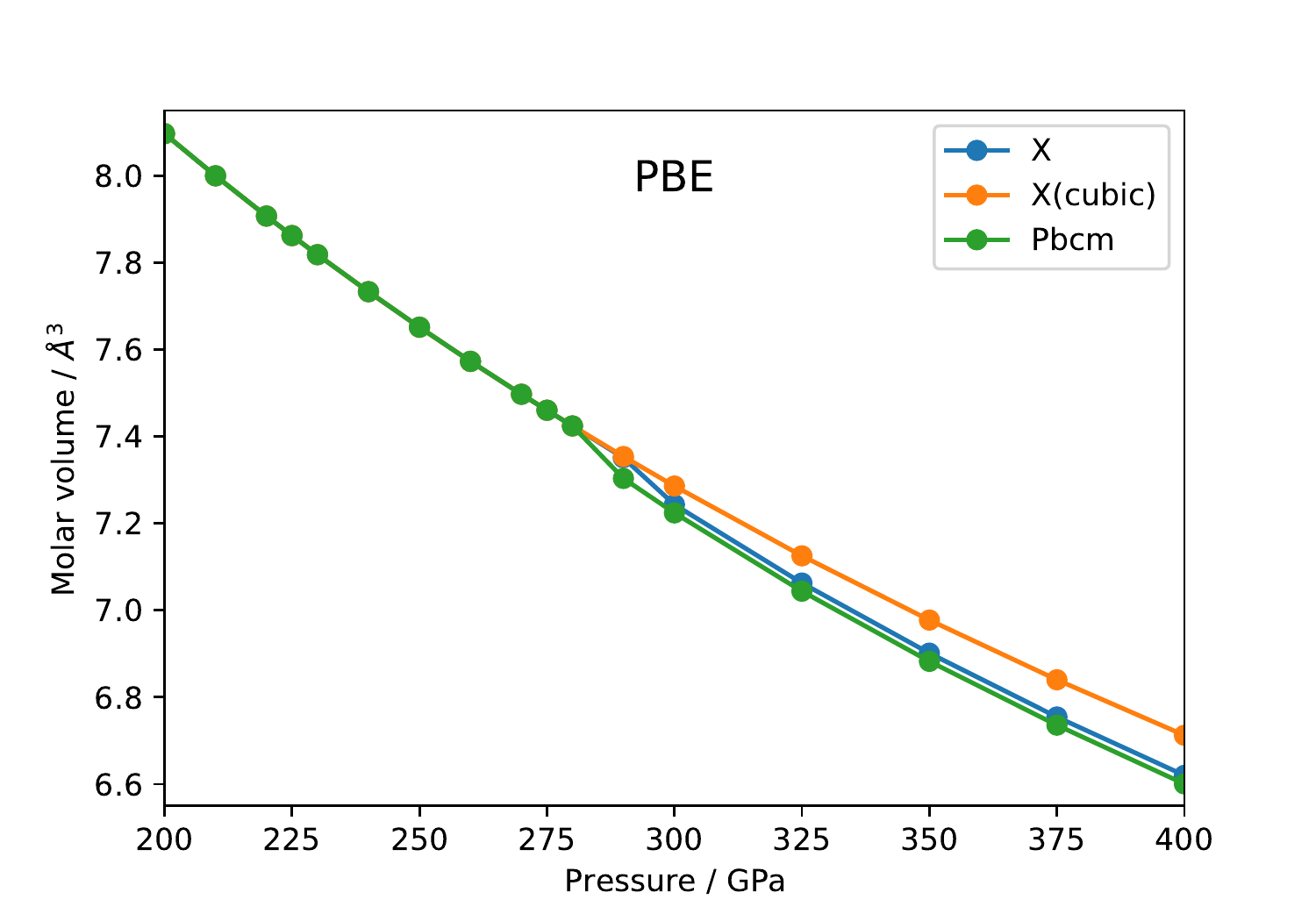}
 \end{tabular}
   \caption{\SI{0}{\kelvin} \figLabelCapt{a} enthalpy, \figLabelCapt{b} potential energy and \figLabelCapt{c} molar volume for different ice phases predicted by the PBE DFT functional.}
   \label{fig:SI-zoomed-curves}
\end{figure}

\begin{figure}[H]
   \centering
  \includegraphics[width=0.45\textwidth]{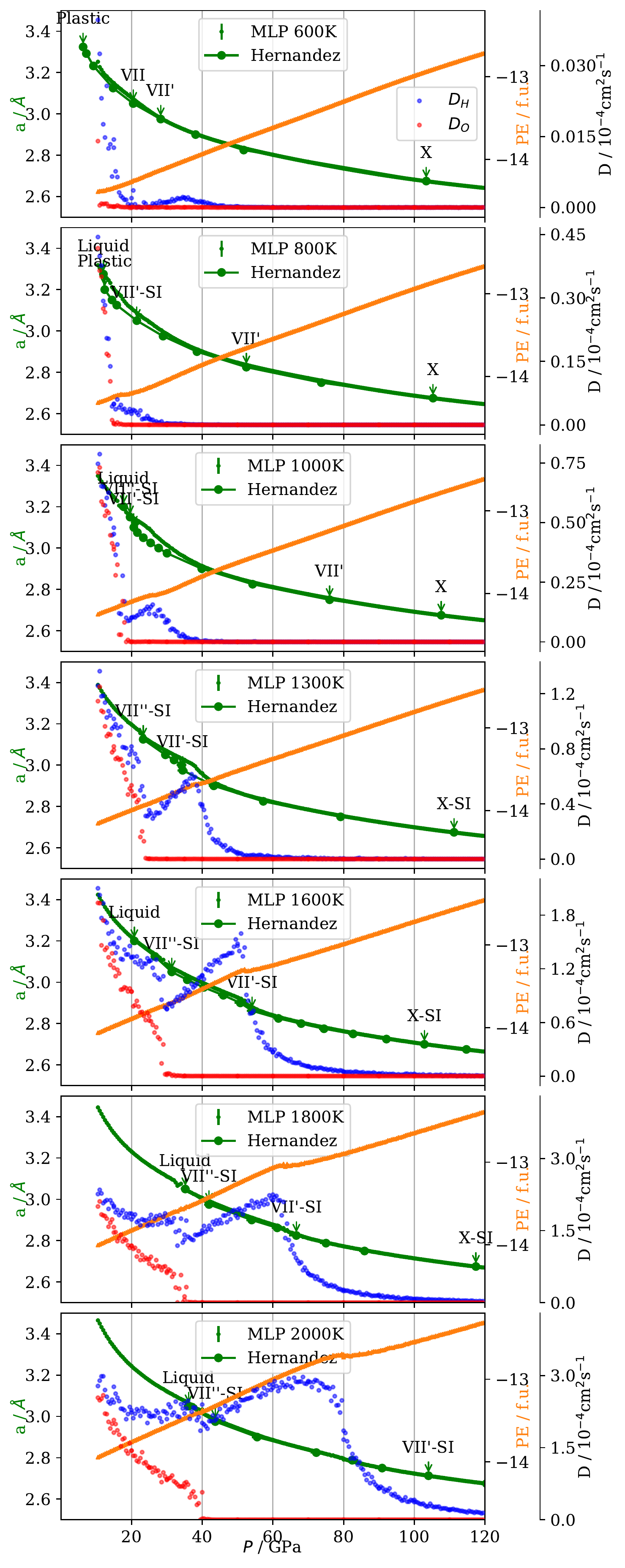}
   \caption{
   Equations of states and diffusivity from MD simulations using the MLP and FPMD. 
    The lattice constant data from FPMD (solid green curves) are from $NVT$ simulations of 128 molecules of Hernandez and Caracas [Ref.~\citenum{Hernandez2018}].
    The onsets of the pressure where different water phases (e.g.~liquid, plastic, VII, \sevenP, \sevenP-SI,
    \sevenPp-SI, X, X-SI) were observed in FPMD simulations are also reproduced from the same reference,
    and these pressures are marked by green arrows.
    The MLP MD data are from $NPT$ simulations using 432 water molecules. 
    The lattice constants are plotted using green dots,
    and the potential energies per molecule are plotted using orange dots.
    The error bars are estimated from the standard error of the mean.
    The red and the blue dots indicate the diffusion coefficient of oxygen and hydrogen atoms, respectively.   }
   \label{fig:X-EOS}
\end{figure}

\begin{figure}[H]
   \centering
   \includegraphics[width=0.95\linewidth]{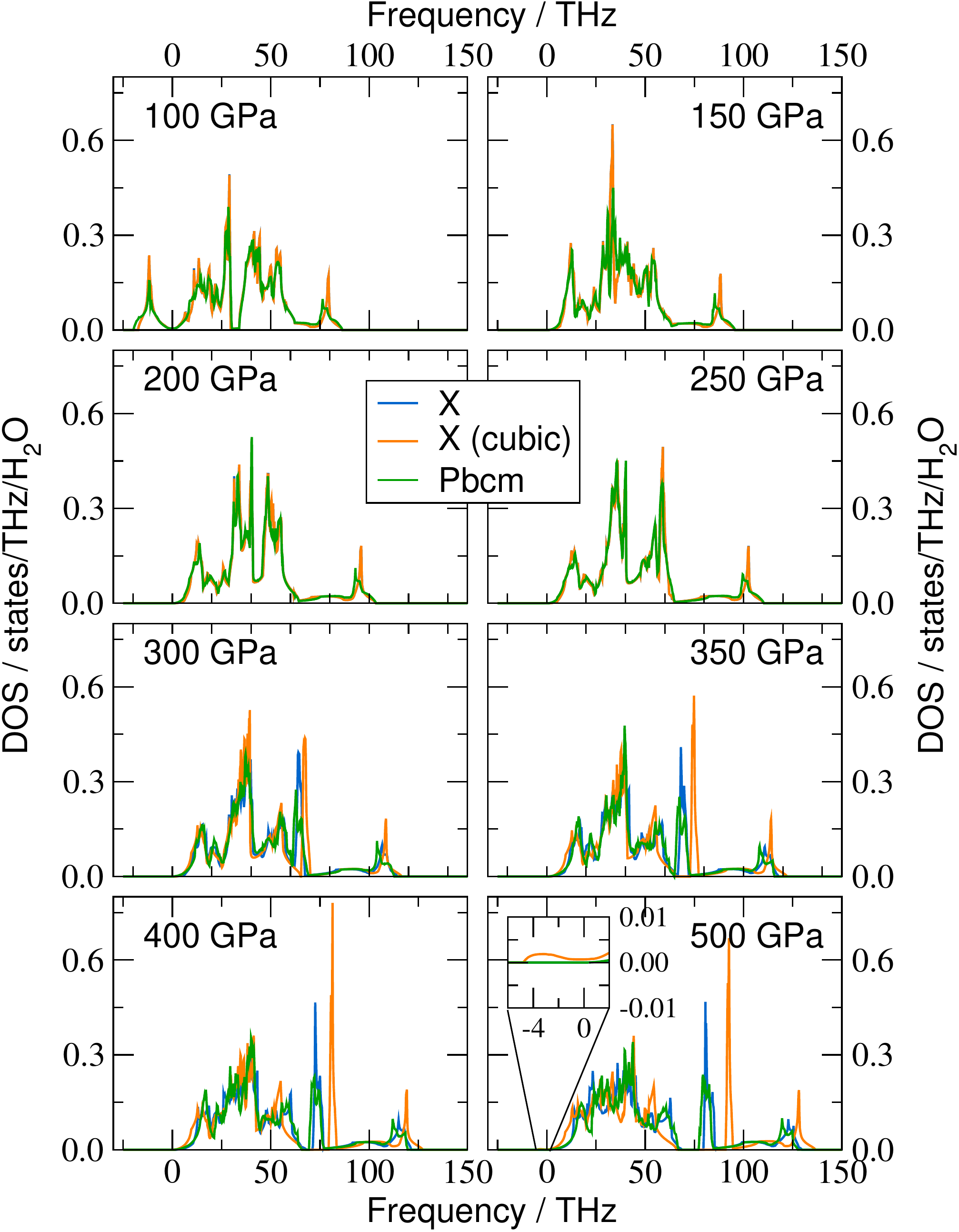}
   \caption{Phonon density of states (DOS) for the three considered structures.} %
   \label{fig:SI-phonon-DOS}
\end{figure}

\begin{figure*}
   \centering
  \includegraphics[width=0.4\textwidth]{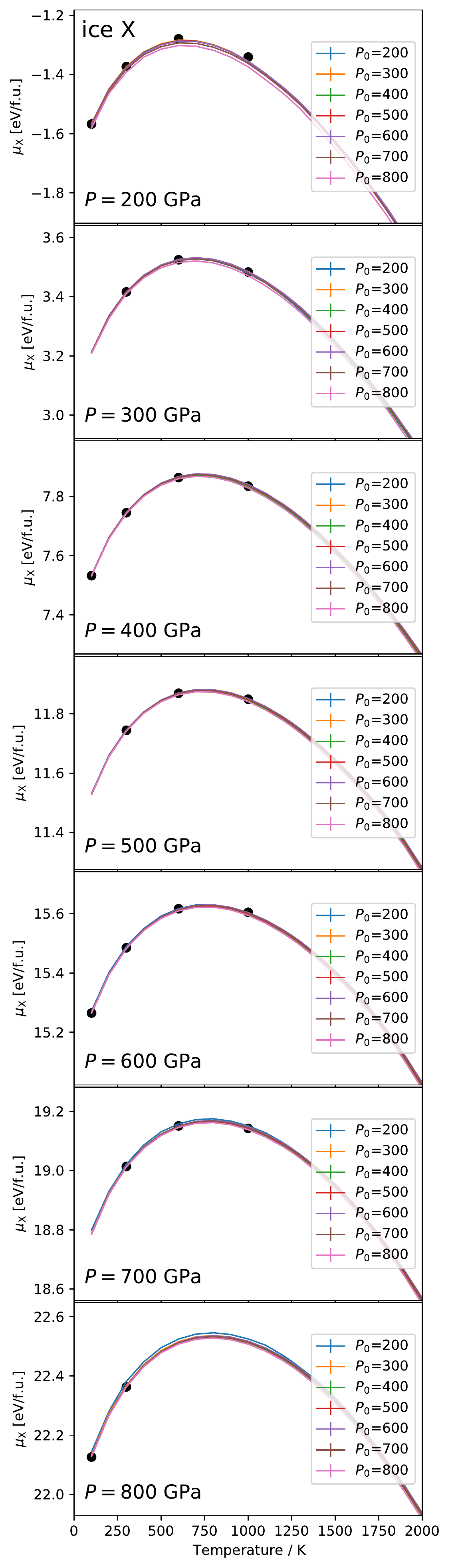}
     \includegraphics[width=0.4\textwidth]{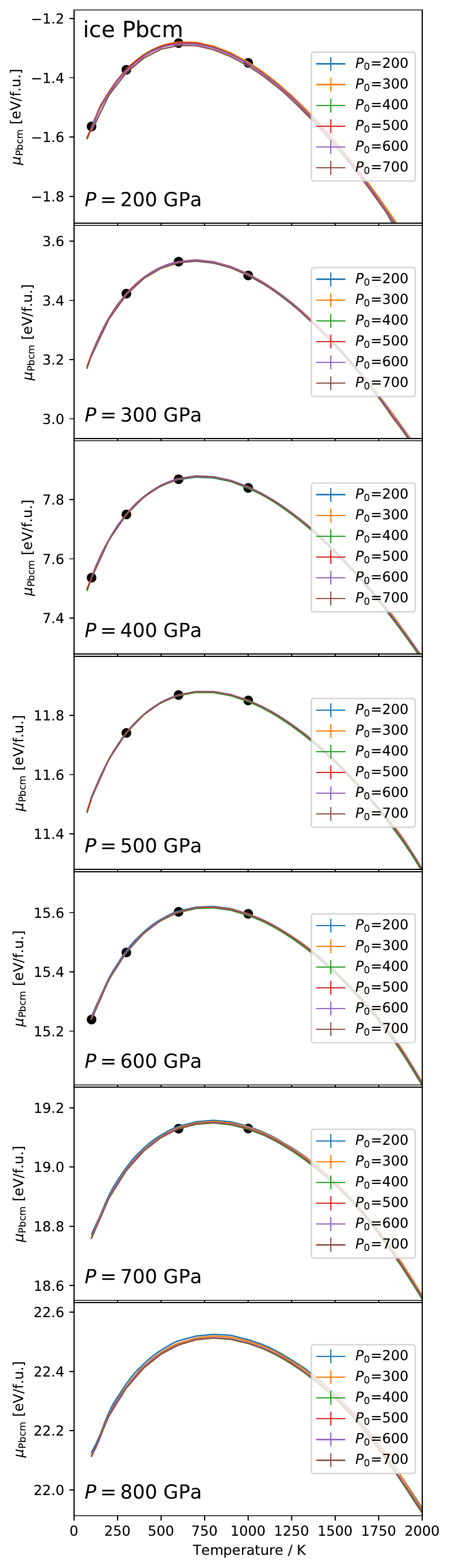}
   \caption{Chemical potentials of X and \Pbcm{} computed along different TI routes.
    The black dots indicate the results obtained from $\lambda$-TI at $P$ and the given temperature.
    The solid lines show the chemical-potential curves obtained from (i) first $\lambda$-TI at $T=\SI{300}{\kelvin}$ or \SI{600}{\kelvin} at pressure $P_0$,
    (ii) integrating along $T$,     and (iii) if $P \neq P_0$, finally integrating along pressure.}
   \label{fig:SI-mu-x-pbcm}
\end{figure*}

\begin{figure}[tbp]
   \centering
   
   \begin{tabular}{@{}p{0.45\textwidth}@{\quad}p{0.45\textwidth}@{}}
   \subfigimg[width=\linewidth]{\vspace*{-15mm}\figLabel{a}}{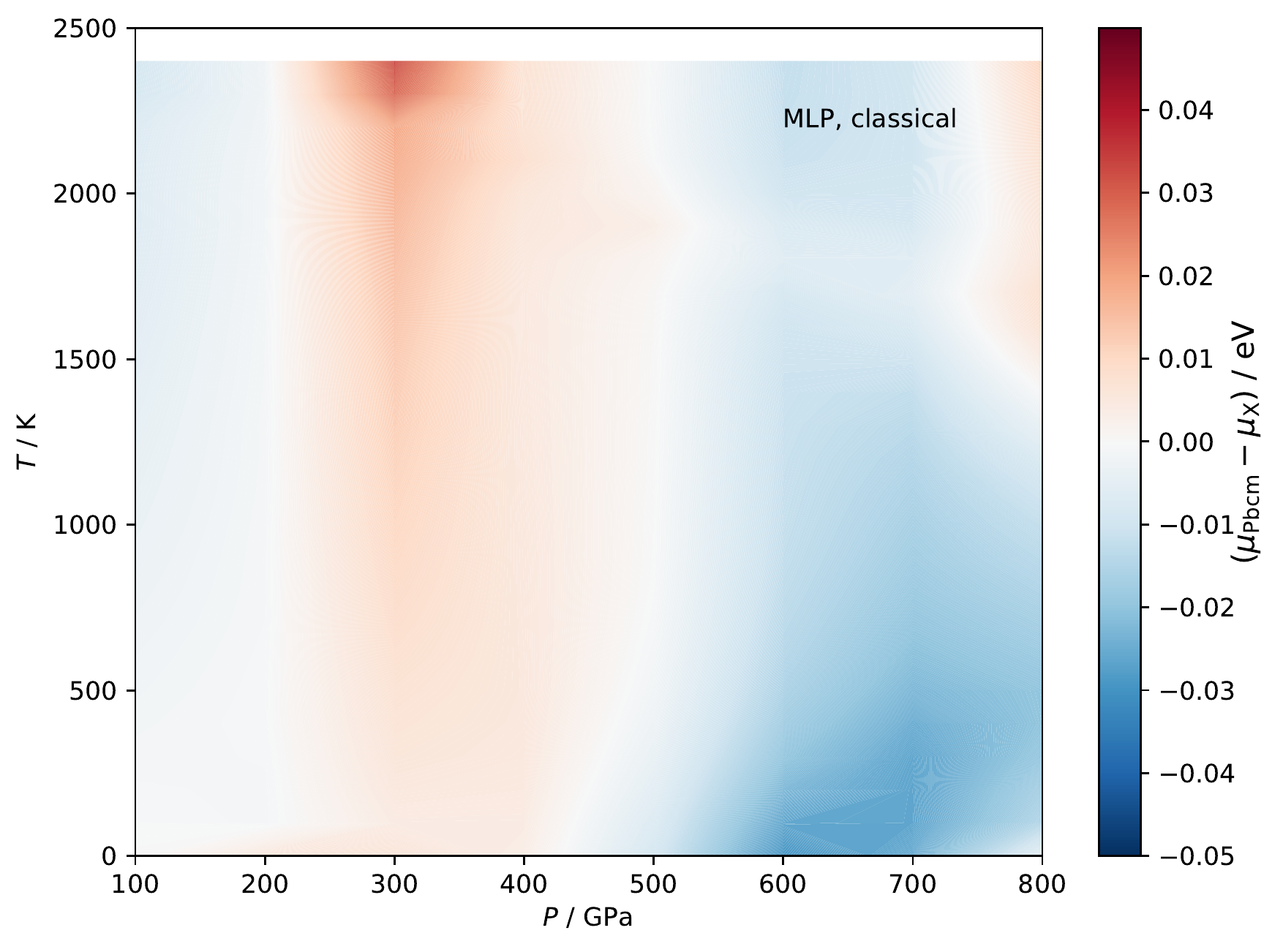} \\
   \subfigimg[width=\linewidth]{\figLabel{b}}{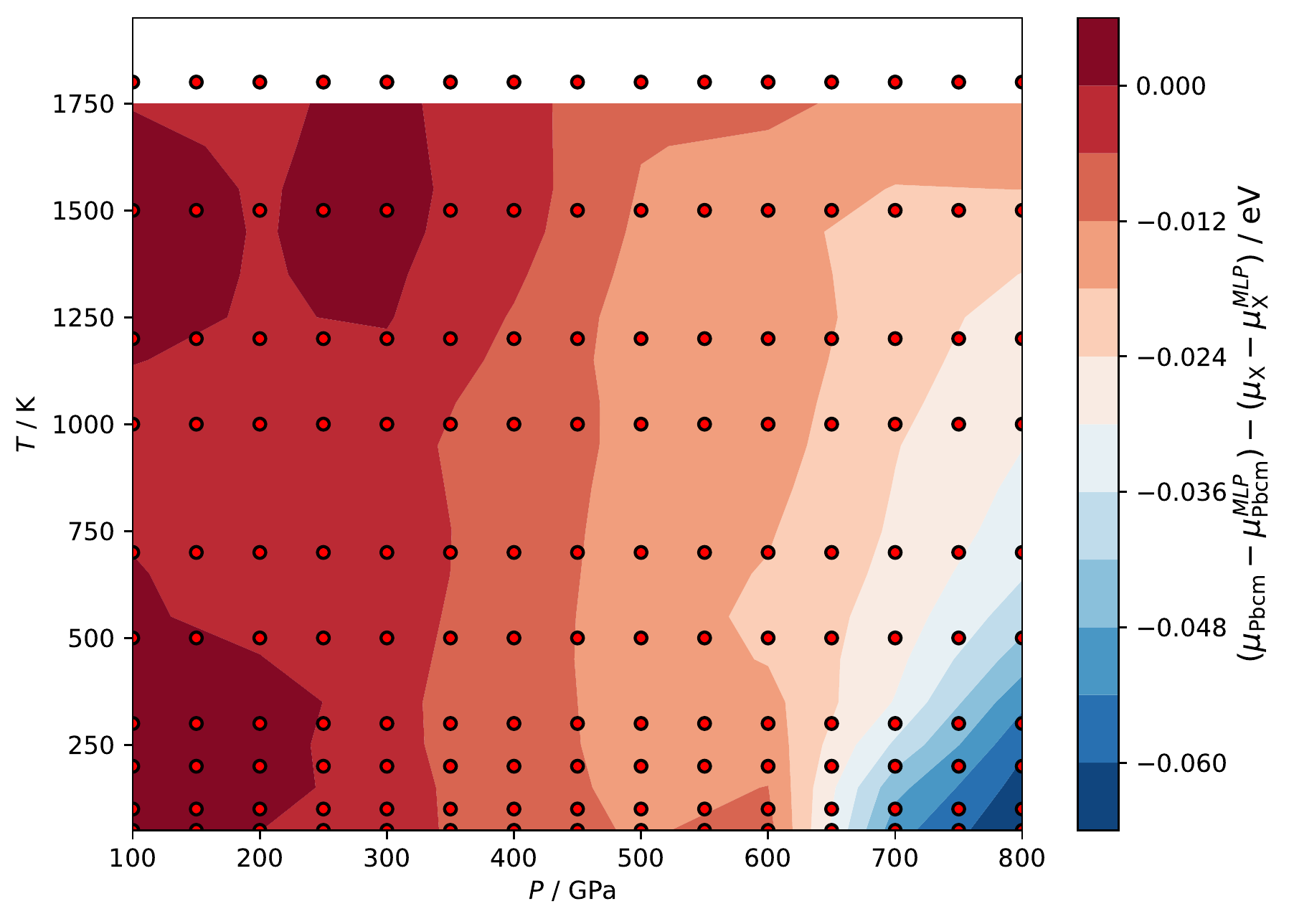}  \\
   \subfigimg[width=\linewidth]{\figLabel{c}}{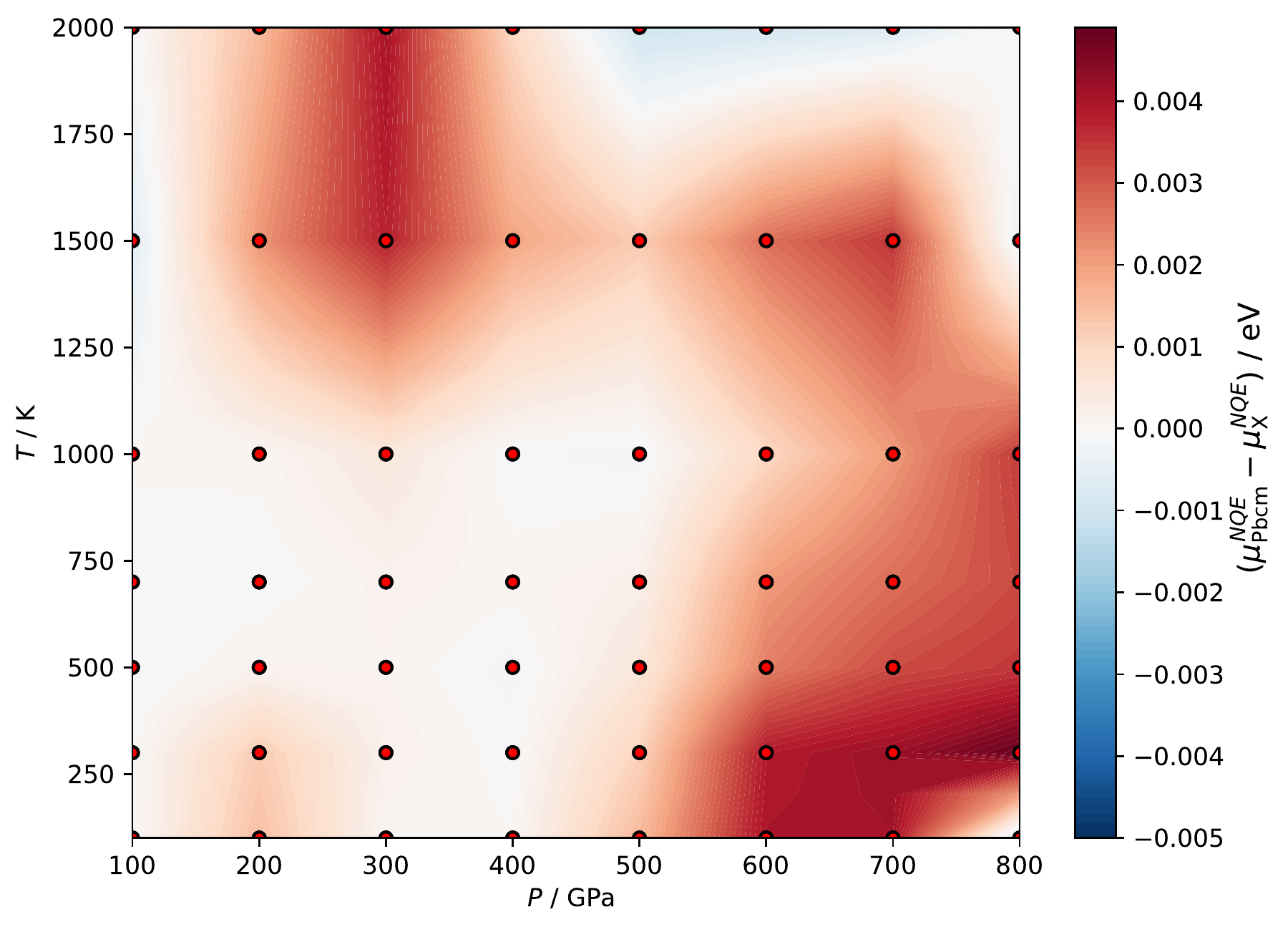}
 \end{tabular}
   \caption{\figLabelCapt{a} The chemical-potential difference between \Pbcm{} and X at the classical nucleus level based on the MLP.
    \figLabelCapt{b} The correction term in the chemical-potential difference from MLP to PBE $\mu-\mu^\mathrm{MLP}$ computed using the free-energy perturbation method.
    \figLabelCapt{c} The contribution of NQEs to the chemical-potential difference.}
   \label{fig:SI-mu-x-pbcm-all}
\end{figure}

\end{document}